\documentclass[conference]{IEEEtran}

\usepackage{cite}
\usepackage{amsmath,amssymb,amsfonts}
\usepackage{textcomp}

\usepackage{graphicx}
\usepackage{amsmath,bm}
\usepackage{xcolor}

\usepackage{algorithm}
\usepackage[noend]{algpseudocode}

\algrenewcommand\alglinenumber[1]{#1}

\usepackage{multirow}

\usepackage{enumitem}
\setlist[itemize]{leftmargin=*, topsep=0pt, itemsep=0pt, parsep=0pt, partopsep=0pt}
\setlist[enumerate]{leftmargin=*, topsep=0pt, itemsep=0pt, parsep=0pt, partopsep=0pt}

\usepackage{amsthm}
\newtheoremstyle{def_style}
  {0pt}          
  {0pt}          
  {}          
  {}          
  {\bfseries} 
  {.}         
  {.5em}      
  {}          

\theoremstyle{def_style}

\theoremstyle{def_style}
\newtheorem*{prob}{Problem Statement}

\begin{document}

\title{Temporal Treasure Hunt: Content-based Time Series Retrieval System for Discovering Insights}

\author{\IEEEauthorblockN{Chin-Chia Michael Yeh, Huiyuan Chen, Xin Dai, Yan Zheng, Yujie Fan, Vivian Lai, \\Junpeng Wang, Audrey Der$^*$, Zhongfang Zhuang, Liang Wang, Wei Zhang}
\IEEEauthorblockA{\textit{Visa Research}, \textit{UC Riverside}$^*$ \\
\{miyeh,hchen,xidai,yazheng,yufan,viv.lai,junpenwa,zzhuang,liawang,wzhan\}@visa.com, ader003@ucr.edu}
}

\maketitle

\begin{abstract}
Time series data is ubiquitous across various domains such as finance, healthcare, and manufacturing, but their properties can vary significantly depending on the domain they originate from.
The ability to perform \textit{Content-based Time Series Retrieval} (CTSR) is crucial for identifying unknown time series examples.
However, existing CTSR works typically focus on retrieving time series from a single domain database, which can be inadequate if the user does not know the source of the query time series. 
This limitation motivates us to investigate the CTSR problem in a scenario where the database contains time series from multiple domains.
To facilitate this investigation, we introduce a CTSR benchmark dataset that comprises time series data from a variety of domains, such as motion, power demand, and traffic. 
This dataset is sourced from a publicly available time series classification dataset archive, making it easily accessible to researchers in the field.
We compare several popular methods for modeling and retrieving time series data using this benchmark dataset. 
Additionally, we propose a novel distance learning model that outperforms the existing methods.
Overall, our study highlights the importance of addressing the CTSR problem across multiple domains and provides a useful benchmark dataset for future research.
\end{abstract}



\section{Introduction}
Time series data is ubiquitous across various domains as almost all systems change over time~\cite{rakthanmanon2012searching}. 
Financial indices, electrocardiograms, and pressure sensors on manufacturing machines provide time-varying measurements of distinct systems, among many other examples. 
This leads institutions to commonly maintain collections of time series data from multiple sources, requiring an effective \textit{Content-based Time Series Retrieval} (CTSR) system for users to efficiently solve their time series-related problems.

\begin{figure}[t]
\centerline{
\includegraphics[width=0.9\linewidth]{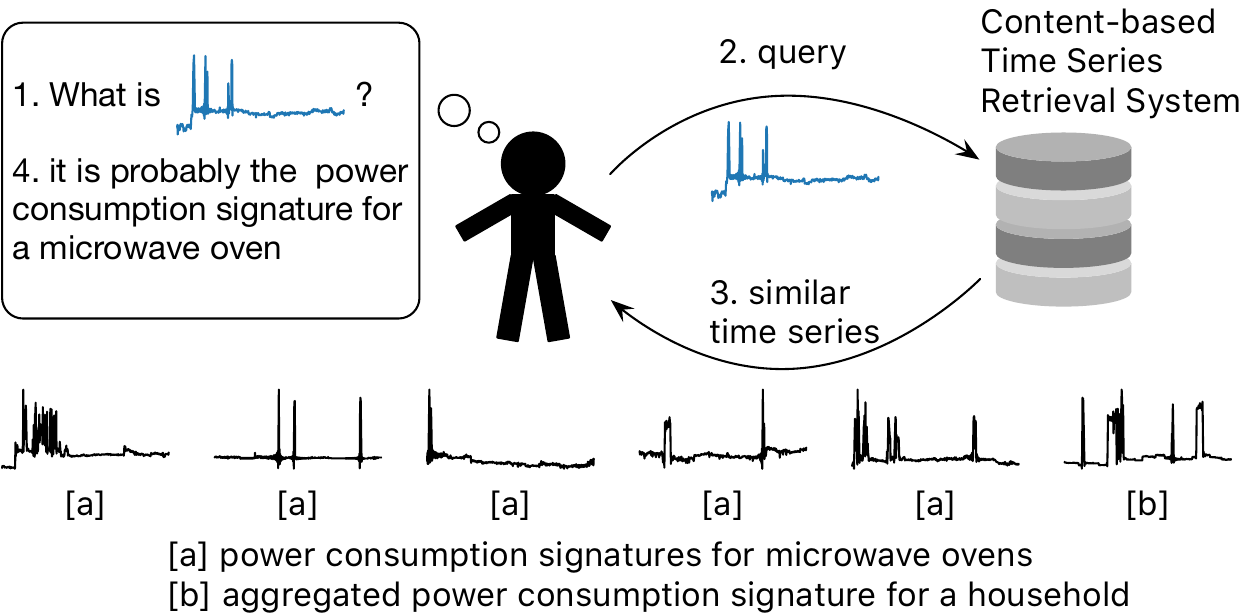}
}
\caption{
The use case for a \textit{Content-based Time Series Retrieval} (CTSR) system is illustrated in the example using the actual outputs generated by the proposed \textit{Residual Network 2D} (RN2D) model.
}
\label{fig:ctsr_exp}
\end{figure}

To understand how a CTSR system can assist users, let us consider a use case as depicted in Fig.~\ref{fig:ctsr_exp}.
Assume that the user acquires a time series from an unknown source and wishes to determine what type of time series it is.
To solve this problem, the user queries a CTSR system with the time series, and the system provides a ranked list of similar time series with relevant metadata.
For example, in our use case, the system returns a list where five out of six similar time series are power consumption signatures for microwave ovens.
The user can then deduce that the unknown time series is likely a power consumption signature for a microwave oven.
Therefore, the CTSR system helps the user to identify potential sources for the query time series.


The primary challenge in addressing the CTSR problem is the distinct characteristics exhibited by time series data from different sources.
An effective CTSR approach must be capable of distinguishing between diverse concepts in various domains.
Nevertheless, existing works in the field have mainly focused on retrieval time series from single domain databases, with the query and database originating from the same domain~\cite{muhammad2010multi,esling2012time,rakthanmanon2012searching,song2018deep,zhu2020deep}, thus ignoring the challenge of retrieving time series from multiple domains.

This work contributes to the study of the CTSR problem in the following ways:
\begin{itemize}
    \item We present a detailed procedure for testing and comparing CTSR systems.
    \item We benchmark six popular baseline methods that are based on commonly used models for time series classification or general sequence modeling.
    \item We propose a novel method called the Residual Network $2D$ (RN2D), which demonstrates superior performance compared to the baseline methods with statistical significance.
\end{itemize}

\section{Method}

In this section, we will first present the baseline methods. 
Next, we will introduce our proposed method and discuss its benefits by contrasting it with the baselines. The CTSR problem is formulated as follows: 
\begin{prob}
\label{prob:ctsr}
Given a set of time series~$\mathcal{X}=[\mathbf{x}_1,\cdots,\mathbf{x}_n]$ and a query time series~$\mathbf{q}$, we aim to obtain a relevance score function~$f(\cdot,\cdot)$ such that $f(\mathbf{x}_i, \mathbf{q}) > f(\mathbf{x}_j, \mathbf{q})$ if $\mathbf{x}_i$ is more relevant to $\mathbf{q}$ compared to $\mathbf{x}_j$.
\end{prob}
\noindent Note that the scoring function $f(\cdot,\cdot)$ can be either a predefined similarity/distance function or a trainable function that is optimized using the time series in $\mathcal{X}$ and their associated metadata.

The baseline methods we considered are:
\begin{itemize}
\item \textbf{Euclidean Distance (ED)}: 
We compute the Euclidean distance between the query time series and each time series in the collection, and then order the collection based on the computed distances. 
This is the simplest method for solving the CTSR problem.
\item \textbf{Dynamic Time Warping (DTW)}: 
This method is similar to the ED baseline, but it uses the Dynamic Time Warping (DTW) distance instead. 
The DTW distance is considered a simple but effective baseline for time series classification problems~\cite{rakthanmanon2012searching,bagnall2017great,dau2019ucr}, and therefore it is important to include this method in our CTSR benchmark.
\item \textbf{Long Short-Term Memory Network (LSTM)}: 
LSTM is one of the most popular types of Recurrent Neural Networks (RNNs) used for modeling sequential data~\cite{hochreiter1997long,lim2021time,zhou2021informer}. 
In this study, we optimize the LSTM models based on the Siamese network architecture~\cite{chicco2021siamese}. 
The input to the Siamese network is composed of two time series. 
We begin by processing each time series using a $1D$ convolutional layer to extract local features. 
We then feed the output of the convolutional layer to a bi-directional LSTM model to obtain a hidden representation. 
The hidden representation of the last time step is fed to a linear layer to get the final representation of the input time series. 
We compute the relevance score between the two input time series by applying the Euclidean distance function to the two final representations (i.e., one from each input time series).
\item \textbf{Gated Recurrent Unit Network (GRU)}: 
The GRU is another popular RNN architecture for modeling sequential data~\cite{cho2014properties,lim2021time,zhou2021informer}.
In our experiments, we optimize the GRU models using the same Siamese network architecture as the LSTM model.
The only difference is that we replace the bi-directional LSTM model with a bi-directional GRU model.
\item \textbf{Transformer (TF)}: 
The transformer has emerged as a popular alternative to RNNs for sequence modeling~\cite{vaswani2017attention,li2019enhancing,zhou2021informer,lim2021time,chen2022denoising,yeh2023egonetwork}.
To learn the hidden representation of the input time series, we use the transformer encoder introduced in~\cite{vaswani2017attention}.
In contrast to the previous methods that employed RNN-based encoders/sub-networks within the Siamese network, we use transformer-based encoders/sub-networks, where the transformer encoders are used to encode the input time series and the relevance score between the two time series is defined as the Euclidean distance between their final transformer outputs.
\item \textbf{Residual Network 1D (RN1D)}: 
This network design is inspired by the success of the residual network in computer vision~\cite{he2016deep,wang2017time}.
Instead of $2D$ convolutional layers, it uses $1D$ convolutional layers~\cite{he2016deep,wang2017time}.
The design was first proposed in~\cite{wang2017time} and was shown to be one of the strongest classifiers for time series classification in extensive evaluations conducted by~\cite{ismail2019deep}.
We optimize this model in a Siamese network architecture~\cite{chicco2021siamese} in the same way as the LSTM method.
\end{itemize}

Both the ED and DTW methods are parameter-free and do not require a training phase. 
However, the DTW method is considered more effective for time series data, as it considers all alignments between the input time series~\cite{rakthanmanon2012searching}. 
The computation of the DTW distance can be described by a two-stage process. 
In the first stage, a pairwise distance matrix~$D \in \mathbb{R}^{w \times h}$ is computed from the input time series $\mathbf{a}=[a_1, \cdots, a_w]$ (where $w$ is the length of $\mathbf{a}$) and $\mathbf{b}=[b_1, \cdots, b_h]$ (where $h$ is the length of $\mathbf{b}$) as $D[i, j]=|a_i-b_j|$. 
In the second stage, a fixed recursion function is applied to $D$, where each element in $D$ is updated as $D[i,j] \gets D[i,j] + \textsc{min}(D[i-1, j], D[i, j-1], D[i-1, j-1])$. 
As a result, the DTW method can be understood as applying a predefined function to the pairwise distance matrix between the input time series.



The other four baseline methods utilize the distance learning framework of the Siamese network. 
They utilize high-capacity\footnote{We use the term \textit{capacity} to describe the \textit{expressiveness}~\cite{lu2017expressive} of a model.} neural network models (LSTM, GRU, TF, and RN1D) to learn the hidden representation of the input time series, which is then used to compute the distance between two time series. 
However, since the neural network models process the two input time series independently, it is unlikely for the model to learn the distance based on different alignments between a pair of time series.

\subsection{Residual Network 2D}
\begin{figure}[ht]
\centerline{
\includegraphics[width=0.99\linewidth]{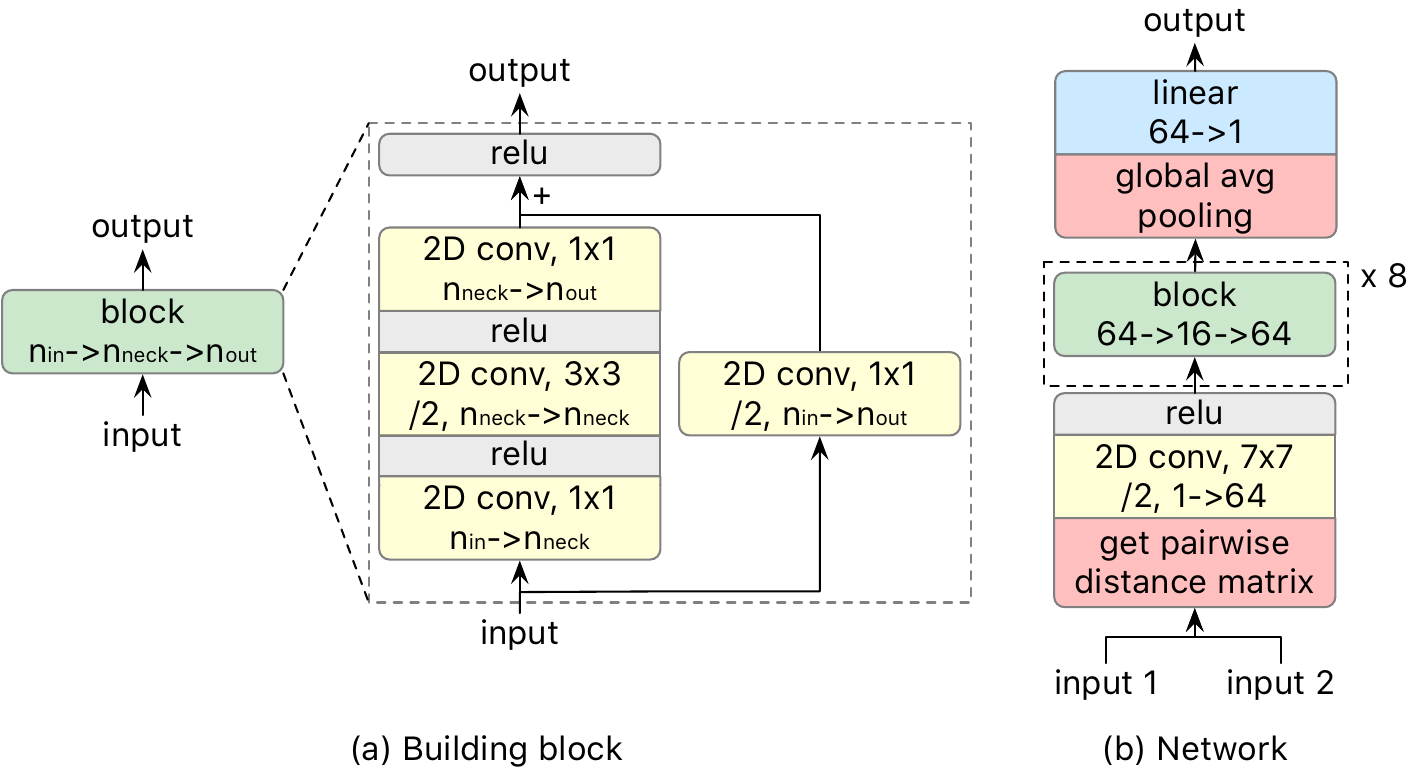}
}
\caption{
The building block and network designs for the proposed \textit{Residual Network 2D} (RN2D) model are shown in Fig.~\ref{fig:resnet_2d}.a and Fig.~\ref{fig:resnet_2d}.b, respectively.
}
\label{fig:resnet_2d}
\end{figure}

To leverage the benefits of both alignment information and neural networks, we propose the \textit{Residual Network 2D} (RN2D) model, as illustrated in Fig.~\ref{fig:resnet_2d}. 
The RN2D model capitalizes on the rich alignment information of the pair-wise distance matrix, similar to the DTW method. 
However, instead of using a fixed function, the proposed method employs a high-capacity neural network to model the distance computation. 
In other words, the RN2D model utilizes an expressive model like the four neural network baselines. 
Fig.~\ref{fig:resnet_2d}.a and Fig.~\ref{fig:resnet_2d}.b depict the building block and overall model design, respectively.

The RN2D model is designed with inspiration from deep residual networks used in computer vision~\cite{he2016deep}. 
We employ the bottleneck building block design from~\cite{he2016deep} as shown in Fig.~\ref{fig:resnet_2d}.a.
Given the input dimension $n_{in}$, bottleneck dimension $n_{neck}$, and output dimension $n_{out}$, we first project the input tensor $X_{\text{in}} \in \mathbb{R}^{w \times h \times n_{in}}$ to $\mathbb{R}^{w \times h \times n_{neck}}$ space using a $1 \times 1$ convolutional layer. 
We then apply a ReLU layer to the intermediate representation, followed by a $3 \times 3$ convolutional layer with a stride of two, which reduces the size further to $\mathbb{R}^{w/2 \times h/2 \times n_{neck}}$.
Next, we use another ReLU layer to process the intermediate representation, followed by a $1 \times 1$ convolutional layer to project it to $\mathbb{R}^{w/2 \times h/2 \times n_{out}}$ space. 
We refer to the output of this layer as $X_{\text{out}}$.
Since the size of $X_{\text{in}}$ does not match that of $X_{\text{out}}$, we first process $X_{\text{in}}$ with a $1 \times 1$ convolutional layer before adding it to $X_{\text{out}}$ for the skip connection. 
Finally, we process the merged representation with a ReLU and exit the building block.
If the input is in $\mathbb{R}^{w \times h \times n_{in}}$ space, the output will be in $\mathbb{R}^{w/2 \times h/2 \times n_{out}}$ space.

Our overall network design is also inspired by the design proposed in~\cite{he2016deep}, as depicted in Fig.~\ref{fig:resnet_2d}.b.
Given two input time series $\mathbf{a}=[a_1, \cdots, a_w]$ and $\mathbf{b}=[b_1, \cdots, b_h]$, we first compute the pairwise distance matrix~$D \in \mathbb{R}^{w \times h}$ similar to the DTW method. 
The $(i,j)$-th position of~$D$ is computed as $D[i, j]=|a_i-b_j|$.
Before applying the convolutional layer, we reshape $D$ to $w \times h \times 1$ by adding an extra dimension.
Next, a $7 \times 7$ convolutional layer with a step size of two projects $D$ to $\mathbb{R}^{w/2 \times h/2 \times 64}$ space. 
After a ReLU layer, the intermediate representation passes through eight building blocks with a $64{\to}16{\to}64$ setting. 
Subsequently, we apply a global average pooling layer to reduce the spatial dimension, and the output of this layer is a 64-dimensional vector. 
Lastly, a linear layer projects the vector to a scalar number, which is the relevance score between the two input time series.




\section{Dataset and Experiment Setup}
We create the CTSR benchmark dataset from the UCR Archive~\cite{dau2019ucr}, a collection of 128 time series classification datasets from various domains like motion, power demand, and traffic.
It is one of the most popular dataset archives for benchmarking time series classification algorithms~\cite{bagnall2017great,wang2017time,ismail2019deep}.
We convert the UCR Archive~\cite{dau2019ucr} to a CTSR benchmark dataset following the steps outlined in~\cite{supplementary}.
After the conversion steps, we end up with 136,377 time series in the training set, 17,005 time series in the test set, and 17,005 time series in the validation set.
To evaluate the performance, we calculate standard information retrieval metrics such as precision at $k$ (Prec@$k$), average precision at $k$ (AP@$k$), and normalized discounted cumulative gain at $k$ (NDCG@$k$) for each query.

We utilized the Bayesian Personalized Ranking (BPR) loss~\cite{rendle2009bpr} as the loss function since the CTSR problem is a Learning to Rank problem.
The loss function is defined as follows for a batch of training data $\mathcal{B}=[B_0, \cdots, B_m]$:
$- \sum_{(\mathbf{t}, \mathbf{t}_+, \mathbf{t}_-)\in \mathcal{B}} \log \sigma(f_{\bm{\theta}}(\mathbf{t}, \mathbf{t}_+) - f_{\bm{\theta}}(\mathbf{t}, \mathbf{t}_-))$
where $m$ is the batch size, $\sigma(\cdot)$ is the sigmoid function, and $f_{\bm{\theta}}(\cdot,\cdot)$ is the model.
Each sample in the training data is a tuple that consists of the query (or anchor), the positive, and the negative time series, i.e., $B_i=(\mathbf{t}, \mathbf{t}_+, \mathbf{t}_-)$.
We used the average NDCG@10 on validation data to select the model for testing.
Further details regarding the source code, hyper-parameter settings, and other information can be found in~\cite{supplementary}.

\section{Experiment Result}
\label{sec:experimentresult}
The performance of the different methods is computed by averaging the performance measurements at $k=10$ across the 17,005 test queries, and the results are presented in Table~\ref{tab:exp_result}.
The table provides an easy way to compare the performance of different methods using various metrics. 
To evaluate the significance of differences between methods, we perform two-sample t-tests with $\alpha=0.05$.

\begin{table}[ht]
\centering
\caption{The proposed method outperforms the others in all performance measurements.
}
\label{tab:exp_result}
\begin{tabular}{l||ccc}
Method & PREC@10 & AP@10 & NDCG@10 \\ \hline \hline
ED & 0.7316 & 0.7655 & 0.7499 \\
DTW & 0.8562 & 0.8792 & 0.8693 \\
LSTM & 0.9205 & 0.9277 & 0.9260 \\
GRU & 0.9108 & 0.9185 & 0.9166 \\
TF & 0.9146 & 0.9212 & 0.9199 \\
RN1D & 0.9086 & 0.9164 & 0.9146 \\ \hline
RN2D & \textbf{0.9266} & \textbf{0.9342} & \textbf{0.9325}
\end{tabular}
\end{table}

We first discuss the performance of the two non-neural network baselines, namely ED and DTW. 
The DTW method significantly outperforms the ED method in all three performance measurements, indicating that the use of alignment information is beneficial for the CTSR problem. 
A similar observation has also been made for the time series classification problem~\cite{bagnall2017great}.
When considering the four neural network baselines (LSTM, GRU, TF, and RN1D), they all significantly outperform the DTW method, highlighting the importance of using high-capacity models for the CTSR problem. 
One possible reason for this is that the CTSR dataset comprises time series from diverse domains~\cite{dau2019ucr}. 
Higher capacity models are required to capture the diverse patterns within the data. 
Among the four methods, LSTM performs significantly better than the second-best method in all three performance measurements. 
The proposed method, RN2D, which is a high-capacity model that utilises alignment information, significantly outperforms all other methods based on the results of t-tests.

\begin{figure}[ht]
\centerline{
\includegraphics[width=0.99\linewidth]{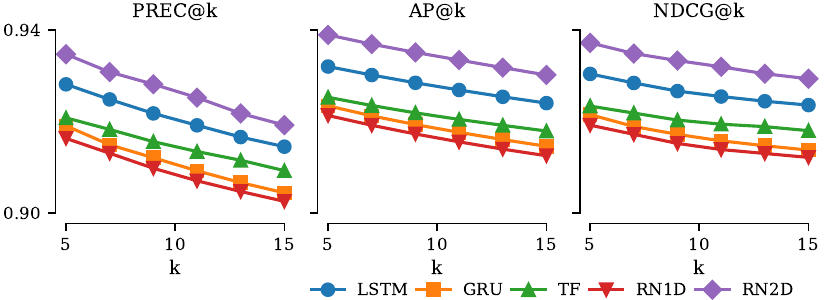}
}
\caption{
The proposed method outperforms the others with different settings of $k$ for each performance measure.
}
\label{fig:k_plot}
\end{figure}

Fig.~\ref{fig:k_plot} presents the performance measurements of different methods at various values of $k$ (ranging from 5 to 15).
This analysis allows us to ensure that the superior performance of the proposed method is not limited to a specific value of $k$.
The performance of ED and DTW is notably lower than the other methods, so we have excluded these two methods in the figures for improved readability.
The RN2D method consistently achieves the highest performance across different values of $k$ for all three performance metrics.
The order of methods from the best to the worst is: RN2D, LSTM, TF, GRU, and RN1D, which aligns with the findings in Table~\ref{tab:exp_result}.

\begin{figure}[t]
\centerline{
\includegraphics[width=0.85\linewidth]{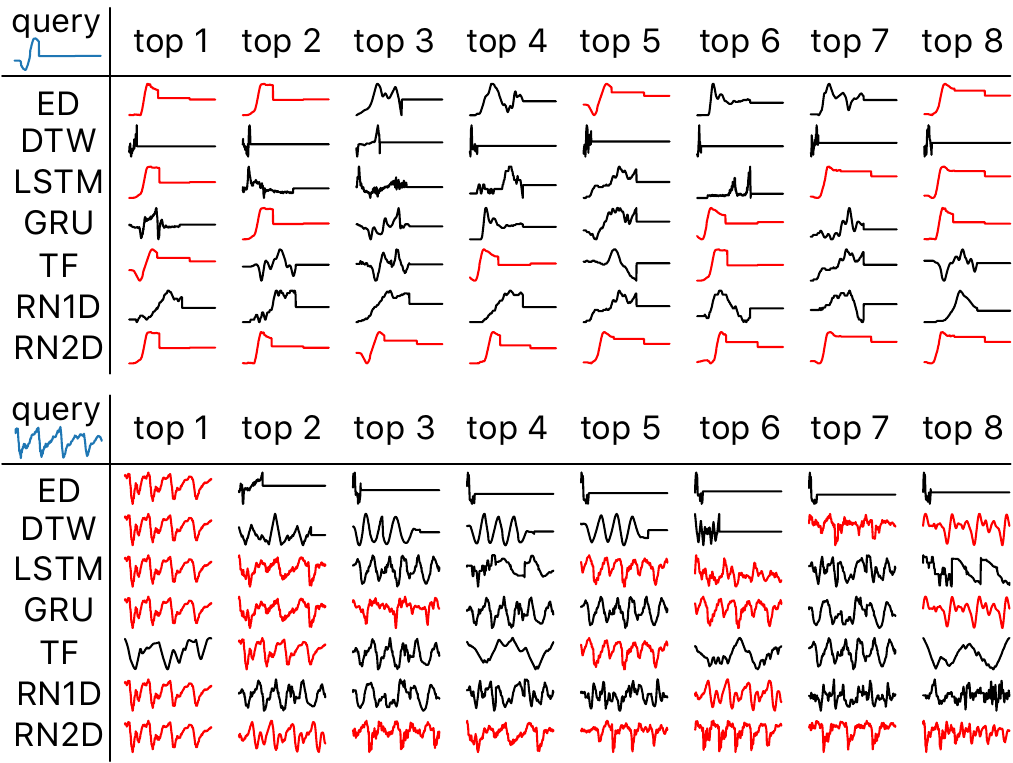}
}
\caption{
The top eight items of a given query time series were retrieved by different methods.
The retrieved time series is colored \textcolor{red}{red} if it is relevant and black if it is irrelevant.
}
\label{fig:query_exp}
\end{figure}

We present the results of the top eight retrieved time series using different methods in Fig.~\ref{fig:query_exp}, where we sample two queries with different complexities from the test dataset. 
The simpler query contains a single cycle of a pattern, while the complex query contains periodic signals, which typically require shift invariant distance measures~\cite{paparrizos2015k} to correctly find the relevant items.
This figure shows the CTSR problem is challenging, as even irrelevant time series are visually similar to the query.

We make the following observations from the retrieved time series of different methods. 
The ED method struggles with the more complex query because it is unable to align the query to relevant time series. 
The DTW method performs better than the ED method on the complex query, but the freedom of alignment hurts its performance on the simple query. 
The four neural network baselines (i.e., LSTM, GRU, TF, and RN1D) work better than both ED and DTW methods when considering both queries. 
However, none of these baselines are able to outperform the proposed RN2D method, which reliably retrieves relevant items. 
Therefore, this figure demonstrates that the proposed RN2D method outperforms other methods by effectively leveraging both the high-complexity model and rich alignment information, enabling it to successfully handle both simple and complex queries.

\section{Conclusion}
In this paper, we investigated the Content-based Time Series Retrieval (CTSR) problem and tested six baseline methods (ED, DTW, LSTM, GRU, TF, and RN1D), and proposed a novel method, called RN2D. 
Our proposed method outperformed the baselines with statistical significance. 
For future work, we plan to focus on improving the computational efficiency of the proposed model by exploring techniques discussed in~\cite{wang2015learning,wang2017survey,yeh2022embedding,chen2022tinykg,andoni2022learning,yeh2023efficient}.
We could also improve the effectiveness of the CTSR system by considering pretraining techniques~\cite{ma2023survey,yeh2023toward}.

\bibliographystyle{IEEEtran}
\bibliography{section/reference.bib}

\end{document}